\documentclass[12pt]{article}
\usepackage{epsfig,latexsym,amssymb,amsmath}
\usepackage{color}
\usepackage{graphicx}
\usepackage[linktocpage,pagebackref]{hyperref}

\begin{document}

\begin{center}
{\large \bf Entropy Product Formula for spinning BTZ Black Hole}
\end{center}

\vskip 5mm

\begin{center}
{\Large{Parthapratim Pradhan\footnote{E-mail: pppradhan77@gmail.com}}}
\end{center}

\vskip  0.5 cm

{\centerline{\it Department of Physics}}
{\centerline{\it Vivekananda Satavarshiki Mahavidyalaya}}
{\centerline{\it (Affiliated to Vidyasagar University)}}
{\centerline{\it Manikpara, West Midnapur}}
{\centerline{\it West Bengal~721513, India}}

\vskip 1cm

\begin{abstract}
We investigate the thermodynamic properties of inner and outer horizons in the background of spinning 
BTZ(Ba\~{n}ados,Teitelboim and Zanelli) black hole. 
We compute the \emph{horizon radii product, the entropy product, the surface temperature product, 
the Komar energy product and the specific heat product} for both the horizons. 
We observe that the entropy product is \emph{universal}(mass-independent), whereas the surface temperature product, 
Komar energy product and specific heat product are \emph{not universal} because they all depends on mass parameter.
We also show that the \emph{First law} of black hole thermodynamics and \emph {Smarr-Gibbs-Duhem }
relations hold for inner horizon as well as outer horizon.  The Christodoulou-Ruffini mass formula is derived 
for both the horizons. We further study the \emph{stability} of such black hole by computing the specific heat for 
both the horizons. It has been observed  that under certain condition the black hole possesses \emph{second 
order phase transition}.
\end{abstract}


\section{Introduction}
There has recently been intense interest in the physics of thermodynamic product formulae in recent 
years due to the seminal work by Ansorg et. al.\cite{ah09} from the gravity site and due to the 
seminal work by Cveti\v{c} et. al. \cite{cgp11} from the string theory site. Of particular interest 
are relations that are independent of mass so called ADM(Arnowitt-Deser-Misner) mass of the back 
ground spacetime and then these relations are said to be ``universal'' in black hole(BH) 
physics\cite{sd12,cr12,pp14,pp15,xu1}. 

They are novel in the sense that they involve the thermodynamic quantities defined at multi-horizons, i.e. 
the Cauchy(inner) horizon and event(outer) horizons of spherically symmetric charged, axisymmetric charged 
and axisymmetric non-charged black hole. For example, let us consider first spherically symmetric charged 
BH, i.e. Reissner Nordstr{\o}m(RN) BH, the mass-independent relations could be found from the 
entropy product of both the horizons(${\mathcal H}^{\pm}$):
\begin{eqnarray}
{\cal S}_{+} {\cal S}_{-} &=& (\pi Q^2)^2 ~.\label{arn}
\end{eqnarray}

For axisymmetric non charged BH, i.e. for Kerr BH, the mass independent relations could be found from 
\begin{eqnarray}
{\cal S}_{+} {\cal S}_{-} &=& (2\pi J)^2 ~.\label{prK1}
\end{eqnarray}

Finally for charged axisymmetric BH\cite{ah09}, these relations are
\begin{eqnarray}
{\cal S}_{+} {\cal S}_{-} &=& (2\pi J)^2 + (\pi Q^2)^2~.\label{prKN1}
\end{eqnarray} 

In the above formulae, the common point is that they all are independent of the mass, so-called the
ADM(Arnowitt-Deser-Misner) mass of the background spacetime. Thus they all are universal quantities in
this sense. Now if we incorporate the BPS(Bogomol'nyi-Prasad-Sommerfeld) states, the entropy product 
formula of ${\mathcal H}^{\pm}$ should be 
\begin{eqnarray}
{\cal S}_{+} {\cal S}_{-}  &=& (2\pi)^2 \left(\sqrt{N_{1}}\pm\sqrt{N_{2}}\right)
= N , \,\, N\in {\mathbb{N}}, N_{1}\in {\mathbb{N}}, N_{2} \in {\mathbb{N}} ~.\label{ss}
\end{eqnarray}
where the integers $N_{1}$ and $N_{2}$ may be viewed as the excitation numbers of the left and right 
moving modes of a weakly-coupled two-dimensional conformal field theory(CFT). $N_{1}$ and $N_{2}$ 
depend explicitly on all the black hole parameters. It also implies that the product of 
${\mathcal H}^{\pm}$ could be expressed in terms of the underlying 
CFT which would be interpreted in terms of a level matching condition. That means the product of 
entropy of ${\mathcal H}^{\pm}$ is an integer quantity \cite{cgp11}.

The entropy product formula of inner horizon and outer horizons could be used to determine whether 
the corresponding Bekenstein-Hawking entropy\cite{bk73} may be written as a Cardy formula, therefore providing 
some evidence for a CFT description of the corresponding microstates\cite{cr12,sd12}. This encourages 
to study the properties of the inner horizon thermodynamics in contrast with outer horizon thermodynamics.

It is a well known fact that certain non-extremal BH possesses inner horizon or Cauchy horizon(CH) in 
addition to the outer horizon or event horizon. Thus there might be a relevance of inner horizon in BH 
thermodynamics to understanding the microscopic nature of inner BH entropy in comparison with the outer 
BH entropy. It is also true that CH is a blue-shift region whereas event horizon is a red-shift region by 
its own characteristics .Furthermore the CH is highly unstable due to the exterior perturbation\cite{sc83}.
Despite the above features, the CH horizon is playing an important role in BH  thermodynamics.

Thus in this note we wish to study various thermodynamic products of rotating BTZ BH \cite{btz92}. We have 
considered both the inner horizon and outer horizons to  understanding the microscopic nature of BH entropy. 
We compute various thermodynamic products like horizon radii product, BH entropy product, surface temperature 
product and Komar energy product. We also study the First law of black hole thermodynamics and Smarr-Gibbs-Duhem
relations for inner horizon as well as outer horizon.  The Christodoulou-Ruffini mass formula is also 
computed for both the horizons. By computing the specific heat, we also analyze the stability of 
such BHs.

The plan of the paper is as follows. In Sec. 2, we described the basic properties of the BTZ BH and 
computed various thermodynamic products.  Finally, we conclude our discussions in Sec. 3.

\section{BTZ  BH:}
The geometric structure in 2+1 dimensions, with an Einstein gravity and a negative cosmological 
constant is described by the BTZ metric\cite{btz92,cai97} reads 
\begin{eqnarray}
ds^2 &= & -\left(\frac{r^2}{\ell^2}+\frac{J^2}{4r^2}-{\cal M} \right) dt^2 
+\frac{dr^2}{\left(\frac{r^2}{\ell^2}+\frac{J^2}{4r^2}-{\cal M} \right)}
+r^2\left(-\frac{J}{2r^2} dt+d\phi\right)^2  ~.\label{btzm}
\end{eqnarray}
where ${\cal M}$ and $J$ are two integration constants and represent the ADM mass, and 
the angular momentum of the BHs. $-\Lambda=\frac{1}{\ell^2}$ denotes the cosmological constant.
Here we have used the $8G=1=c=\hbar=k$. In the limit $J=0$, we obtain the usual behavior of 
a static BTZ BH.

The BH horizons are\cite{btz92}:
\begin{equation} 
r_{\pm}= \sqrt{\frac{{\cal M} \ell^2}{2}\left(1\pm \sqrt{1-\frac{J^2}{{\cal M}^2 \ell^2}} \right)}
 .~\label{hor}
\end{equation}
As is $r_{+}$ corresponds to event horizon and  $r_{-}$ corresponds to Cauchy horizon.
As long as
\begin{eqnarray}
1-\frac{J^2}{{\cal M}^2 \ell^2} \geq 0 ~.\label{ineq}
\end{eqnarray}
then the BTZ metric  describes a BH, otherwise it has a naked  singularity. When $r_{+}=r_{-}$ or 
$1-\frac{J^2}{{\cal M}^2 \ell^2}=0$, we find an extremal BTZ BH.

The product and sum  of horizon radii becomes
\begin{equation}
r_{+}^2 r_{-}^2 = \frac{J^2\ell^2}{4}\,\,\, \mbox{and} \,\, r_{+}^2+ r_{-}^2 = {\cal M} \ell^2
 ~.\label{hrp}
\end{equation}
 It is clearly evident that the product is independent of mass whereas the sum depends on mass 
parameter.

The entropy of this BH is given by
\begin{eqnarray}
{\cal S}_{\pm} &= & 4\pi r_{\pm}  ~.\label{epb}
\end{eqnarray}
Their product and sum yields 
\begin{eqnarray}
{\cal S}_+ {\cal S}_- &=& 8 \pi^2 J \ell \,\,\, \mbox{and} \,\, {\cal S}_{+}^2+ {\cal S}_{-}^2 = 16 \pi^2 {\cal M} \ell^2
~.\label{epsz}
\end{eqnarray}
It is remarkable that the entropy product of BTZ BH is independent of mass but their sum depends on the 
mass parameter. 

For completeness, we further compute the entropy minus and entropy division:
\begin{eqnarray}
{\cal S}_{\pm}^2- {\cal S}_{\mp}^2 &=& \pm 16 \pi^2 {\cal M} \ell^2 \sqrt{1-\frac{J^2}{{\cal M}^2 \ell^2}} 
~.\label{enzpm}
\end{eqnarray}
and  
\begin{eqnarray}
\frac{{\cal S}_{+}}{{\cal S}_{-}} &=& \frac{r_{+}}{r_{-}} ~.\label{entzd}
\end{eqnarray}

Again, the sum of entropy inverse is found to be 
\begin{eqnarray}
\frac{1}{{\cal S}_{+}^2}+\frac{1}{{\cal S}_{-}^2} &=& \frac{{\cal M}}{4\pi^2 J^2}
 ~.\label{etpiz}
\end{eqnarray}
and the minus of entropy inverse is computed to be 
\begin{eqnarray}
\frac{1}{{\cal S}_{\pm}^2}-\frac{1}{{\cal S}_{\mp}^2} &=& \mp \frac{{\cal M}}{4 \pi^2 J^2} 
\sqrt{1-\frac{J^2}{{\cal M}^2 \ell^2}} ~.\label{entiz}
\end{eqnarray}
It suggests that they are all mass dependent relations.

The Hawking\cite{bcw73} temperature of ${\cal H}^{\pm}$ reads off
\begin{eqnarray}
T_{\pm} &=& \frac{r_{\pm}^2-r_{\mp}^2}{2\pi \ell^2 r_{\pm}}=\pm \frac{{\cal M}\sqrt{1-\frac{J^2}{{\cal M}^2 \ell^2}}}
{2\pi  \sqrt{\frac{{\cal M} \ell^2}{2}\left(1\pm \sqrt{1-\frac{J^2}{{\cal M}^2 \ell^2}} \right)}}.~\label{tempz}
\end{eqnarray}
Their product and sum yields
\begin{eqnarray} 
T_{+} T_{-} &=& -\frac{{\cal M}^2}{2\pi^2 J \ell}\left(1-\frac{J^2}{{\cal M}^2 \ell^2}\right) 
\,\,\, \mbox{and} \,\,\,T_{+} +T_{-} = \frac{(r_{+}^2-r_{-}^2)(r_{-}-r_{+})}{\pi J \ell^3}  
.~\label{tempsz}
\end{eqnarray}
It may be noted that surface temperature product and sum both depends on mass thus they are 
not universal in nature. 
It is shown that for BTZ BH:
\begin{eqnarray}
T_{+}{\cal S}_{+}+T_{-}{\cal S}_{-} &=& 0 .~\label{ts}
\end{eqnarray}
which is same as in Kerr BH\cite{ok92}. It is in-fact a mass independent(universal) relation and 
implies that $T_{+}{\cal S}_{+}= -T_{-}{\cal S}_{-}$
should be taken as a criterion whether there is a 2D CFT dual for the BHs in the Einstein gravity and other 
diffeomorphism gravity theories\cite{chen12,xu}. This universal relation also indicates that the left and right 
central charges are equal i.e. $c_{L}=c_{R}$ that means it is holographically dual to 2D CFT.

The angular velocity is computed on both the horizons(${\cal H}^{\pm}$) as
\begin{eqnarray}
\Omega_{\pm} &=& \frac{J}{2r_{\pm}^2} .~\label{zz}
\end{eqnarray}

Their product and sum yields:
\begin{eqnarray}
\Omega_{+}\Omega_{-} &=& \frac{1}{\ell^2}  \,\,\, \mbox{and} \,\,\, \Omega_{+}+\Omega_{-} = \frac{2{\cal M}}{J} 
.~\label{zz1}
\end{eqnarray}
Finally, the Komar\cite{ak59} energy computed at ${\cal H}^{\pm}$ is given by
\begin{eqnarray}
E_{\pm} &=& \frac{4}{\ell^2}(r_{\pm}^2-r_{\mp}^2) .~\label{komz}
\end{eqnarray}
Their product and sum reads off
\begin{eqnarray}
E_{+}E_{-} &=& \frac{16}{\ell^4}(r_{+}^2-r_{-}^2)(r_{-}^2-r_{+}^2) \,\,\, \mbox{and} \,\,\, E_{+}+E_{-} = 0 
.~\label{kpsz}
\end{eqnarray}
which indicates that the product does depend on the mass parameter and thus it is not an universal
quantity.

\subsection{Smarr like formula for BTZ BH on $\mathcal{H}^{\pm}$:}

Smarr\cite{smarr73} first showed  the ADM  mass can be expressed as a function of area, angular momentum and
charge for Kerr-Newman black hole. On the other hand,  Hawking\cite{bcw73} pointed out that the BH area
always increases. Therefore the BH  area is indeed a constant quantity over the ${\cal H}^{\pm}$.
Here, we prove the mass could be expressed in terms of entropy of the both horizons(${\cal H}^{\pm}$).

The entropy of both the horizons for BTZ  BH is given by
\begin{eqnarray}
{\cal S}_{\pm}=4 \pi \sqrt{\frac{{\cal M} \ell^2}{2}\left(1\pm \sqrt{1-\frac{J^2}{{\cal M}^2 \ell^2}} \right)}
\label{z1}
\end{eqnarray}

Alternatively, the ADM mass can be expressed in terms of entropy of the horizons:
\begin{eqnarray}
{\cal M} &=& \frac{{\cal S}_{\pm}^2}{16 \pi^2 \ell^2}+\frac{4\pi^2 J^2 }{ {\cal S}_{\pm}^2}.~\label{z2} 
\end{eqnarray}
After differentiation, we get the mass differential as
\begin{eqnarray}
d{\cal M}=T_{\pm} d{\cal S}_{\pm} +\Omega_{\pm}dJ ~\label{z3}
\end{eqnarray}
where,
\begin{eqnarray}
T_{\pm} &=& \mbox{Hawking temperature for both the horizons}\nonumber\\
        &=& \frac{{\cal S}_{\pm}^4-64\pi^4 \ell^2J^2}{8\pi^2\ell^2 {\cal S}_{\pm}^3}\\
        &=& \frac{r_{\pm}^2-r_{\mp}^2}{2\pi \ell^2 r_{\pm}}\\
        &=& \frac{\partial {\cal M}}{\partial {\cal S}_{\pm}}.\\
\Omega_{\pm} &=& \mbox{Angular velocity for both the horizons }\nonumber\\
             &=&  \frac{J}{2r_{\pm}^2}\\
             &=& \frac{\partial {\cal M}}{\partial J}. ~\label{z4}
\end{eqnarray}
The Smarr-Gibbs-Duhem relation holds for BTZ BH as
\begin{eqnarray}
{\cal M} &=& {T}_{\pm}{\cal S}_{\pm} + J\Omega_{\pm} ~. \label{z5}
\end{eqnarray}
It can be seen that both first law of BH thermodynamics and Smarr-Gibbs-Duhem relation
holds for BTZ BH.
\subsection{ Irreducible mass product for BTZ BH:}
In 1970, Christodoulou\cite{cd70} had shown that the irreducible mass
${\cal M}_{irr}$ of a non-spinning BH is related to the 
surface area ${\cal A}$ by the following relation
\begin{eqnarray}
{\cal M}_{irr, +}^{2} &=& \frac{{\cal A}_{+}}{16\pi}=\frac{{\cal S}_{+}}{4\pi}
~. \label{z6}
\end{eqnarray}
where `$+$' sign indicates for ${\cal H}^{+}$ and now it is established that this 
relation is valid for CH also\cite{pp14}. That implies
\begin{eqnarray}
{\cal M}_{irr, -}^{2} &=& \frac{{\cal A}_{-}}{16\pi}=\frac{{\cal S}_{-}}{4\pi}
~. \label{z7}
\end{eqnarray}
Here, `$-$' indicates for ${\cal H}^{-}$.

Equivalently, the expressions for entropy on ${\cal H}^{\pm}$, in terms of ${\cal M}_{irr, \pm}$ are:
\begin{eqnarray}
{\cal S}_{\pm} &=& 4 \pi ({\cal M}_{irr, \pm})^2 ~. \label{z8}
\end{eqnarray}
Therefore for BTZ BH, the product of the irreducible mass at the horizons ${\cal H}^{\pm}$ are:
\begin{eqnarray}
{\cal M}_{irr, +} {\cal M}_{irr,-}
&=& \frac{J\ell}{2}.~\label{z9}
\end{eqnarray}
The above product is an universal quantity because it does not depend upon the ADM mass
parameter. It should be noted that Eq. (\ref{z2}) can be written as in terms of irreducible mass:
\begin{eqnarray}
{\cal M} &=& \frac{{\cal M}_{irr,\pm}^4}{\ell^2} + \frac{J^2}{4 {\cal M}_{irr,\pm}^4} .~\label{z10}
\end{eqnarray} 
\subsection{Heat Capacity $C_{\pm}$ for BTZ BH on ${\cal H}^{\pm}$:}
The phase transition is an important phenomenon in BH thermodynamics. This can be determine by 
computing the specific heat. Therefore one can define the specific heat on ${\cal H}^{\pm}$:
\begin{eqnarray}
C_{\pm} &=& \frac{\partial{\cal M}}{\partial T_{\pm}} .~\label{c1}
\end{eqnarray}
Finally we get the expression for specific heat:
\begin{eqnarray}
C_{\pm} &=& 4 \pi \frac{\sqrt{1-\frac{J^2}{{\cal M}^2 \ell^2}}}{2-\sqrt{1-\frac{J^2}{{\cal M}^2 \ell^2}}}
\sqrt{\frac{{\cal M} \ell^2}{2}\left(1\pm \sqrt{1-\frac{J^2}{{\cal M}^2 \ell^2}} \right)}. ~\label{c5}
\end{eqnarray}
It should be noted that the  specific  heat $C_{\pm}$ blows up when 
$J^2=-3{\cal M}^2 \ell^2$, in this case the BH undergoes a second order phase transition.
Their product and sum yields:
\begin{eqnarray}
C_{+} C_{-} &=& 8\pi^2 J \ell \frac{1-\frac{J^2}{{\cal M}^2 \ell^2}}
{\left(2-\sqrt{1-\frac{J^2}{{\cal M}^2 \ell^2}}\right)^2}  .~\label{c6}
\end{eqnarray}
and 
\begin{eqnarray}
C_{+}^2 +C_{-}^2 &=& 16\pi^2 \ell^2 {\cal M}  \frac{1-\frac{J^2}{{\cal M}^2 \ell^2}}
{\left(2-\sqrt{1-\frac{J^2}{{\cal M}^2 \ell^2}}\right)^2}  .~\label{c7}
\end{eqnarray}
It indicates that both are mass dependent relations.

It should be noted that in the extremal limit $J^2={\cal M}^2 \ell^2$, the temperature, Komar energy
and specific heat of the BH becomes zero. It also should be noted that in the limit $J=0$, all the above 
results reduce to static BTZ BH. 
\subsection{Entropy Bound for ${\cal H}^{\pm}$:}
Using the above thermodynamic relations, we are now able to derive the entropy bound of both the horizons. 
Since  $r_{+}^2 \geq r_{-}^2$, one can get ${\cal S}_{+}^2 \geq {\cal S}_{-}^2 \geq 0$. 
Then the entropy product gives:
\begin{eqnarray}
{\cal S}_{+}^2  \geq  {\cal S}_{+} {\cal S}_{-}=8\pi^2\ell J \geq {\cal S}_{-}^2 ~.\label{sz1}
\end{eqnarray}
and the entropy sum gives:
\begin{eqnarray}
 16 \pi^2 {\cal M} \ell^2 ={\cal S}_{+}^2+ {\cal S}_{-}^2 \geq {\cal S}_{+}^2 \geq 
\frac{{\cal S}_{+}^2+ {\cal S}_{-}^2}{2}= 8 \pi^2 {\cal M} \ell^2
 ~.\label{sz2}
\end{eqnarray}
Thus the entropy bound for  ${\cal H}^{+}$:
\begin{eqnarray}
 8 \pi^2 {\cal M} \ell^2 \leq {\cal S}_{+}^2 \leq  16 \pi^2 {\cal M} \ell^2
 ~.\label{sz3}
\end{eqnarray}
and  the entropy bound for  ${\cal H}^{-}$:
\begin{eqnarray}
 0 \leq {\cal S}_{-}^2 \leq  8\pi^2\ell J  ~.\label{sz4}
\end{eqnarray}

\subsection{ Irreducible Mass Bound for ${\cal H}^{\pm}$:}
From the entropy bound, we get the irreducible mass bound for BTZ BH.
For ${\cal H}^{+}$:
\begin{eqnarray}
 \left(\frac{{\cal M} \ell^2}{2} \right)^{\frac{1}{4}} \leq {\cal M}_{irr, +} \leq 
 \left({\cal M} \ell ^2\right)^{\frac{1}{4}}   ~.\label{mbz}
\end{eqnarray}
and  for ${\cal H}^{-}$:
\begin{eqnarray}
0 \leq {\cal M}_{irr,-} \leq  \left(\frac{J \ell}{2} \right)^{\frac{1}{4}} ~.\label{mbz1}
\end{eqnarray}
Eq. \ref{mbz} is nothing but the Penrose inequality, which is the first geometric inequality for 
BHs \cite{peni73}.

\section{Discussion}
In this paper, we have studied the thermodynamic features of BTZ BH. We computed
various thermodynamic product formula for this BH. We observed that the BH entropy product formula  
and irreducible mass product formula are universal because they are independent of ADM mass parameter, 
whereas the Hawking temperature product, Komar energy product and specific heat product  formulae 
are not universal quantities because they all are depends on ADM mass parameter. 

We considered the Smarr like relations for the above mentioned BH. We showed the Smarr-Gibbs-Duhem 
relation holds for both the horizons. We also derived the Christodoulou-Ruffini mass formula for 
all the horizons. Based on the termodynamic relations, we derived the entropy bound for all horizons. 
From entropy bound, we computed irreducible mass bound for both the horizons. Finally, we studied 
the stability of such black hole by computing the specific heat for both the horizons. It has been 
shown that in the limit $J^2=-3{\cal M}^2 \ell^2$ the black hole possesses  a second order phase 
transition.  

In summary, these product formulae of ${\cal H}^{\pm}$ in lower dimension gives us further evidence 
for the important role to understanding the BH entropy both interior and exterior at the microscopic 
level.


\end{document}